\documentclass[showpacs,aps,prb,twocolumn]{revtex4}
\usepackage{graphicx}

\def  \bsig   {\mbox{\boldmath$\sigma $}}
\def  \bnu    {\mbox{\boldmath$\nu $}}
\def  \btau   {\mbox{\boldmath$\tau $}}

\begin{document}

\title{Edge scattering of electrons in graphene}

\author{V.~K.~Dugaev}
\affiliation{Department of Physics,
Rzesz\'ow University of Technology, Al.~Powsta\'nc\'ow Warszawy 6,
35-959 Rzesz\'ow, Poland}
\affiliation{Department of Physics and CFIF, Instituto Superior T\'ecnico,
Universidade T\'ecnica de Lisboa, Av. Rovisco Pais, 1049-001 Lisbon, Portugal}

\author{M. I. Katsnelson}
\affiliation{Radboud University Nijmegen, Institute for Molecules and Materials,
Heyendaalseweg 135, 6525 AJ Nijmegen, The Netherlands}

\begin{abstract}
We discuss the contribution of edge scattering to the conductance of graphene
nanoribbons and nanoflakes. Using different possible types of the boundary conditions for the electron wave
function at the edge, we found dependences of the momentum relaxation time
and conductance on the geometric sizes and on the carrier density.
We also consider the case of ballistic nanoribbon and nanodisc, for which
the edge scaterring is the main mechanism of momentum relaxation.
\end{abstract}

\date{\today }
\pacs{73.22.pr, 73.23.-b, 72.10.Fk}

\maketitle

\section{Introduction}

Very unusual transport properties of graphene are mostly related
to the electronic energy structure of low-energy states in this material,
that can be described by the ultrarelativistic Dirac Hamiltonian. \cite{castro09,katsnelson}
The main parameter of this model, electron velocity, does not depend
on the electron energy, and is rather high (about $10^6$~m/s). Besides,
the electron backscattering from impurities is effectively suppressed in graphene (``Klein tunneling'' \cite{katsnelson}).
It results in a rather high mobility of electrons in the graphene bulk despite possible inhomogeneities.
Typically, the bulk electron mean-free-path $\ell $ is just several times smaller than the size of 
graphene flakes $L$ or even comparable with it.
This may lead to important contribution of electron scattering from the edges.
The main parameter, which determines the condition for essential contribution of
the edge scattering, is $\ell /L$. For $\ell /L\ll 1$ the
edge scattering leads to a small correction to the transport coefficient
but in the opposite (ballistic) case, the edge scattering is the main mechanism of
momentum relaxation. Ballistic regime can be experimentally reached for graphene samples. 
\cite{bal1,bal2,bal3}

The effect of electron scattering from the surface has been thoroughly
studied in the past for ordinary metals and semiconductors.
In the framework of kinetic equation approach, the main problem of the
theory is the boundary condition for the electron distribution function at the surface.
It was proposed by Fuchs to use a constant specular factor to formulate the boundary condition.
\cite{fuchs38} It turned out, however, that this approach is too rough to explain numerous experiments.
Besides, such boundary condition is not related to any specific mechanism of the surface
scattering, and quite obviously does not take into account different character
of scattering of electrons incoming under small and large angles to the surface.
The problem has been examined in many papers (see, e.g., Refs.
[\onlinecite{falkovskii70,ustinov80,peschanskii81,katsnelson81}]) and review articles
[\onlinecite{okulov79,falkovsky83}] accounting for different scattering mechanisms
from different kind of defects, including nonmagnetic and magnetic impurities, surface roughness, etc.

Here we discuss the role of edge scattering in graphene.
The essential property of graphene, which makes the results different from
the above mentioned results for conventional metallic systems is the behavior of the wave function
of electron near the edge. Since the low-energy electrons in graphene are described by relativistic  Dirac
model, one cannot assume zero wavefunction at the edge, which is the standard way to introduce the metal surface. 
As a result, the surface scattering vanishes for the sliding electrons with the momentum parallel to the surface,
which is especially essential for the ballistic regime $\ell /L\gg 1$. \cite{okulov79,falkovsky83} 
The boundary conditions for the wave function in graphene turn out to depend on orientation of the edge with respect
to the crystal lattice, on possible edge reconstruction and on the chemical passivation of the edges.
\cite{katsnelson} We will show in this work that it leads, indeed, to an essential difference in the
results from those for conventional metals. 

Several types of the boundary conditions have been proposed.
The so-called Berry-Mondragon \cite{berry87} (or infinite-mass)
boundary conditions are quite universal to describe the confinement of Dirac electrons
in a restricted region as they are not related to the orientation of the boundary. 
They correspond to the single Dirac cone approximation
and therefore are applicable for the case of smooth enough disorder near the edges. 
It seems to be a good approximation for chemically 
functionalized edges since the first-principle calculations show that electronic 
structure is affected at distances much larger 
than the lattice constant. \cite{boukhvalov08} 

The microscopic model for the boundary conditions and the edge states in graphene, which is based on the
real crystallic structure and uses tight-binding approximation, has been considered
in several papers. \cite{mccann04,brey06,akhmerov08,wimmer10} It was found that for the zigzag
boundary, one of the wavefunction components should be necessarily zero at the edge (the other one is zero
at the opposite edge). For the armchair boundary it is important to consider two
nonequivalent Dirac points (i.e., electrons from different valleys), and the boundary conditions
input some phase-dependent relations between the wave functions components of different valleys.
It was shown also that for terminated honeycomb lattice zigzag boundary conditions are robust whereas
the armchair ones are exceptional. \cite{akhmerov08,wimmer10} We will focus therefore on two 
cases, Berry-Mondragon and zigzag edges. 
In both these cases one can neglect intervalley scattering.

However, the situation with graphene nanoribbons and nanoflakes can be more complicated
because of the crystallic reconstruction of the edge, which makes some types of the edges
like, e.g., ``reczag'' reconstruction, energetically more favorable. \cite{koskinen08} The boundary 
conditions for this case has been derived in Ref. [\onlinecite{ostaay11}]. 
In general, they include the intervalley scattering,
which is also relevant for the case of atomically sharp disorder at the edges. 

The plan of the paper is the following. In Section II we consider the general solution 
of the kinetic equation for the graphene nanoribbon, 
in Section III we derive boundary conditions for the kinetic equation for the nanoribbon 
with Berry-Mondragon and zigzag boundary conditions, 
the edge is supposed to be straight line with some defects on it. We will show that the 
surface scattering vanishes for the sliding electrons 
in the case of zigzag boundaries but not for the Berry-Mondragon case. In Section IV we 
calculate the contribution of the edge scattering to 
the conductance of graphene nanoribbon for $\ell /L\ll 1$. In section V the opposite 
limit $\ell /L\gg 1$ is considered. In Section VI we consider 
the scattering by curved edges and in Sections VII and VIII discuss the role of intervalley 
edge scattering. In Section IX we consider the case of 
graphene circular flake (nanodisc) with Berry-Mondragon boundary conditions. We finalize 
with the discussion of the results (Section X) 
and conclusions (Section XI). 

\section{Formulation of the model for graphene nanoribbon}

Let us consider first a narrow graphene ribbon of width $L$ along axis $y$, so that
the graphene edges are located at $x=0$ and $x=L$. We assume first that the ribbon edges
are ideally flat (straight lines).

The energy spectrum of electrons with momentum ${\bf k}$ and
energy $\varepsilon >0$ in the vicinity of $\mathcal{K}$
or $\mathcal{K'}$ Dirac points is $\varepsilon (k)=vk$, where $v$ is a constant,
and energy $\varepsilon $ is measured from the Dirac point.
We assume that graphene is moderately doped, so that the Fermi energy lies at some
$\varepsilon _F>0$ not far from the Dirac point $\varepsilon =0$.

One can justify the use of the standard semiclassical kinetic equation not too close to 
the neutrality point,
namely, for $k_F\ell \gg 1$, where $k_F$ is the Fermi wave vector (or, equivalently, when the 
static conductivity 
$\sigma \gg e^2/h$). \cite{katsnelson,auslender07}
Further we will assume this condition to be fulfilled. 

The kinetic equation for the stationary distribution function of electrons
$f({\bf k},x)=f_0+\delta f$ in an electric field $E$ along axis $y$, with $\delta f$
depending on $x$, reads
\begin{eqnarray}
\label{1}
eE\, \frac{\partial f}{\partial k_y}
+v_x\, \frac{\partial f}{\partial x}
=-\frac{\delta f}{\tau }\, ,
\end{eqnarray}
where $f_0(\varepsilon )$  is the equilibrium distribution function,
$v_i=vk_i/\hbar k$ is the electron velocity, and
$\tau $ is the momentum relaxation time related to the scattering from
impurities or other defects in the graphene bulk.

If the external field $E$ is weak, then we use the linear response approximation
and obtain from Eq. (1)
\begin{eqnarray}
\label{2}
eEv_y\, \frac{\partial f_0}{\partial \varepsilon }
+v_x\, \frac{\partial \delta f}{\partial x}
=-\frac{\delta f}{\tau }\, ,
\end{eqnarray}
where $\varepsilon =v\, (k_x^2+k_y^2)^{1/2}$.
The general solution of Eq.~(2) for $v_x>0$ and for $v_x<0$ can be presented as
\begin{eqnarray}
\label{3}
&&\delta f^>(k_y,x)=-eEv_y\tau \; \frac{\partial f_0}{\partial \varepsilon }
+\mathcal{C}^>(k_y)\, e^{-x/l_x},
\\
&&\delta f^<(k_y,x)=-eEv_y\tau \; \frac{\partial f_0}{\partial \varepsilon }+
\mathcal{C}^<(k_y)\, e^{(x-L)/l_x},
\end{eqnarray}
respectively, where $l_x=|v_x|\tau $, and $\mathcal{C}^>(k_y),\, \mathcal{C}^<(k_y)$ 
are some arbitrary functions, which have to be found from the boundary conditions at the edges.

It should be noted that the solution (3),(4) is not valid in the limit of $\tau \to \infty $.
In such a ballistic limit the functions $\delta f^>$ and $\delta f^<$ do not depend on $x$,
and the electron scattering from the edges should be directly included into the right hand
part of the kinetic equation (1) (see below).

\section{Boundary condition for the distribution function}

At the left edge of the ribbon, $x=0$, one can use the following boundary condition for the
distribution function
\begin{eqnarray}
\label{5}
|v_x|f^>(k_y,0)=|v_x|f^<(k_y,0)
+\int \frac{d^2{\bf k'}}{(2\pi )^2}\;
W_L({\bf k},\bf{k'})
\nonumber \\ \times
\left[ f^<(k'_y,0)-f^>(k_y,0)\right] ,
\hskip0.3cm
\end{eqnarray}
where $W_L({\bf k},{\bf k'})$ is the probability of backscattering at the left edge
from the state ${\bf k}$ to ${\bf k'}$
\begin{eqnarray}
\label{6}
W_L({\bf k},{\bf k'})
=\frac{2\pi N_i}{\hbar }\,
|\left< {\bf k}|V(x,y)|{\bf k'}\right> |^2\,
\delta (\varepsilon _{{\bf k}}-\varepsilon _{{\bf k'}}) ,
\end{eqnarray}
$N_i$ is the linear density of scatterers (defects)
along the graphene edge, and $V(x,y)$ is the potential
of a single scatterer at the edge $x=0$. If there are several different
types of scatterers, the probability $W_L({\bf k,k'})$ is a corresponding sum
of several terms (6).
The boundary condition (5) accounts for the mirror reflection
at the edge and also for reflection from scatterers, which are
assumed to be homogenously distributed along the edge.

Analogously, we can write the boundary condition for the distribution function
at the right edge of the ribbon, $x=L$
\begin{eqnarray}
\label{7}
|v_x|f^<(k_y,L)=|v_x|f^>(k_y,L)
+\int \frac{d^2{\bf k'}}{(2\pi )^2}\;
W_R({\bf k},{\bf k'})
\nonumber \\ \times
\left[ f^>(k'_y,L)-f^<(k_y,L)\right] .
\hskip0.3cm
\end{eqnarray}
For simplicity we assume in the following that the type and distribution of impurities
and defects is the same at both edges, so that
$W_L({\bf k},{\bf k'})=W_R({\bf k},\bf{k'})$.
It means that in average there is the mirror symmetry $k_x\to -k_x$.

\subsection{Berry-Mondragon boundary condition for the wave function}

To calculate the matrix elements of impurity potential $V(x,y)$ in
Eq.~(6) we should use the wave functions $\left| {\bf k}\right> $
describing the electron states near graphene edge.
For this purpose we can write the following Schr\"odinger equation
\begin{eqnarray}
\label{8}
\left( \begin{array}{cc}
\varepsilon -\Delta (x) & v(i\partial _x+ik_y) \\
v(i\partial _x-ik_y) & \varepsilon +\Delta (x)
\end{array}\right)
\left( \begin{array}{c}\varphi \\ \chi \end{array}\right)
=0,
\end{eqnarray}
where $\varphi (x,y)$ and $\chi (x,y)$ are the spinor components of the wave
function $\psi (x,y)$, the gap function
$\Delta (x)=\Delta _0\theta (-x)$, and $\Delta _0>>|\varepsilon |$.
This corresponds to the vacuum at $x<0$ (with a constant large gap $\Delta _0$),
and to the graphene at $x>0$, so that the graphene edge is the line $x=0$.
The boundary condition of this type has been introduced by Berry and
Mondragon.\cite{berry87}

Using Eq.~(8) we find that at $x<0$, $\varphi =Ae^{\kappa _x x+ik_yy}$
and $\chi =Be^{\kappa _x x+ik_yy}$,
whereas at $x>0$, $\varphi =De^{ik_x x+ik_yy}$ and $\chi =Fe^{ik_x x+ik_yy}$.
Substituting this to Eq. (8) we find for $x<0$ (vacuum)
\begin{eqnarray}
\label{9}
(\varepsilon -\Delta _0)A+iv(\kappa _x+k_y)B=0,
\\
iv(\kappa _x-k_y)A+(\varepsilon +\Delta _0)B=0,
\end{eqnarray}
and from the condition of zero determinant of the set of linear equations (9),(10) we obtain
$\kappa _x=\frac1{v}(\Delta _0^2-\varepsilon ^2+v^2k_y^2)^{1/2}\simeq \Delta _0/v$.
Correspondingly, from (9) and (10) follows $B\simeq -iA$.
Due to the continuity of wavefunction at $x=0$ we also obtain $F=-iD=-iA$.

Thus, the wavefunction obeying Berry-Mondragon boundary conditions, near the graphene edge, $x>0$, is
\begin{eqnarray}
\label{11}
\psi _{\bf k}(x,y)=Ae^{i{\bf k}\cdot {\bf r}}
\left( \begin{array}{c} 1 \\ -i \end{array}\right) ,
\end{eqnarray}
and the components of wavevector ${\bf k}$
are related by $v(k_x^2+k_y^2)^{1/2}=\varepsilon $.

We assume that the potential $V(x,y)$, corresponding to a single impurity or defect at the graphene edge,
is short ranged in $x$-direction and has a
characteristic range $a$ in $y$-direction (i.e., along the edge), so that electron scattering
with rather strong $k_y$-momentum
transfer, $|k_y-k'_y|>1/a$, is effectively suppressed. It corresponds to assumption that
the Fourier transform of $y$-dependent random potential does not have wavevector components
with $|k_y-k'_y|>1/a$. Such a model can be used to describe different character of
the edge scattering of electrons incoming under different angles (diffusive
for large angles and nearly specular for small angles).\cite{falkovsky83}
Hence, one can take $\left< {\bf k}|V(x,y)|{\bf k'}\right> \simeq V_0 e^{-(k_y-k'_y)^2a^2}$,
where $V_0$ is a constant. Note that it does not matter, in which sublattice A or B of graphene is located
the impurity with potential $V(x,y)$.

Then the boundary condition (5) can be written as
\begin{eqnarray}
\label{12}
|v_x|f^>(k_y,0)=|v_x|f^<(k_y,0)
+\frac{2\pi N_iV_0^2}{\hbar }
\int \frac{d^2{\bf k'}}{(2\pi )^2}\hskip0.5cm
\nonumber \\ \times
e^{-2(k_y-k'_y)^2a^2}
\delta (\varepsilon _{{\bf k}}-\varepsilon _{{\bf k'}})
\left[ f^<(k'_y,0)-f^>(k_y,0)\right] . \hskip0.2cm
\end{eqnarray}
We can use
\begin{eqnarray}
\label{13}
\delta (\varepsilon _{{\bf k}}-\varepsilon _{{\bf k'}})
=\frac{k\, \delta (k'_x-k'_{x0})}{vk'_x}\, ,
\end{eqnarray}
where $k'_{x0}=(k^2-{k'_y}^2)^{1/2}$.
Then we get from Eq. (12)
\begin{eqnarray}
\label{14}
|v_x|f^>(k_y,0)=|v_x|\, f^<(k_y,0)
+\frac{N_iV_0^2k}{2\pi \hbar v}\hskip1.5cm
\nonumber \\ \times
\int _{-k}^k dk'_y\; e^{-2(k_y-k'_y)^2a^2}
\frac{\left[ f^<(k'_y,0)-f^>(k_y,0)\right] }{(k^2-{k'_y}^2)^{1/2}}\; ,
\end{eqnarray}

Assuming that the scattering from impurities at the edge $x=0$  is weak we 
can substitute $f^>(k_y,0)$ by $f^<(k_y,0)$ in the right-hand part of (14),
and we finally present the boundary
condition for the distrubution function at $x=0$ as
\begin{eqnarray}
\label{15}
|v_x|f^>(k_y,0)=|v_x|\, f^<(k_y,0)
+\frac{N_iV_0^2k}{2\pi \hbar v}\hskip1.5cm
\nonumber \\ \times
\int _{-k}^k dk'_y\; e^{-2(k_y-k'_y)^2a^2}
\frac{\left[ f^<(k'_y,0)-f^<(k_y,0)\right] }{(k^2-{k'_y}^2)^{1/2}}\; .
\end{eqnarray}

Correspondingly, the second boundary condition for the distribution function
at $x=L$ acquires the following form
\begin{eqnarray}
\label{16}
|v_x|f^<(k_y,L)=|v_x|f^>(k_y,L)
+\frac{N_iV_0^2k}{2\pi \hbar v}\hskip1.5cm
\nonumber \\ \times
\int _{-k}^k dk'_y\; e^{-2(k_y-k'_y)^2a^2}
\frac{\left[ f^>(k'_y,L)-f^>(k_y,L)\right] }{(k^2-{k'_y}^2)^{1/2}}\; .
\end{eqnarray}

Substituting Eqs. (3),(4) into Eqs.~(15) and (16) we find the solution for the functions
$\mathcal{C }_{BM}^>(k_y)$ and $\mathcal{C}_{BM}^<(k_y)$ for the Berry-Mondragon boundary
\begin{eqnarray}
\label{17}
\mathcal{C}^>_{BM}(k_y)=\mathcal{C}^<_{BM}(k_y)=
\frac{eE\tau N_iV_0^2}{2\pi \hbar ^2|v_x|\, (1-e^{-L/l_x})}
\left(-\frac{\partial f_0}{\partial \varepsilon }\right)
\nonumber \\
\times \int _{-k_F-k_y}^{k_F-k_y}
\frac{e^{-2q^2a^2}qdq}{[k_F^2-(k_y+q)^2]^{1/2}}\; .
\hskip0.5cm
\end{eqnarray}
This solution is valid for weak disorder at the edge.

\subsection{Zigzag boundary condition for the wave function}

One can also consider ``zigzag'' boundary condition for the wavefunction at the left edge, $x=0$,
as $\varphi (0)=0$. Its status has been discussed above in the Introduction.
Then the wave function at $x>0$ (i.e., in graphene near the edge) has the form
\begin{eqnarray}
\label{18}
\psi _{\bf k}(x,y)\simeq Ae^{ik_yy}\left(
\begin{array}{c} \sin k_xx \\
-\frac{ik_x}{k}\cos k_xx +\frac{ik_y}{k}\sin k_xx
\end{array}
\right)
\end{eqnarray}
Now the matrix element of impurity potential $V(x,y)$ strongly localized in
sublattice A reads
\begin{eqnarray}
\label{19}
\left< {\bf k}|\hat{V}^{(A)}|{\bf k'}\right>
=|A|^2\int d^2{\bf r}\,
\sin (k_xx)\, \sin (k'_xx)\hskip0.5cm
\nonumber \\ \times
e^{-i(k_y-k'_y)y}\, V(x,y)
\simeq V_Ak_xk'_x e^{-(k_y-k'_y)^2a^2},
\end{eqnarray}
where $V_A$ is a constant.

Analogously, we find for impurity potential localized in sublattice B
\begin{eqnarray}
\label{20}
\left< {\bf k}|\hat{V}^{(B)}|{\bf k'}\right>
\simeq \frac{|A|^2k_xk'_x}{k^2}\int d^2{\bf r}\,
\cos (k_xx)\, \cos (k'_xx)\hskip0.5cm
\nonumber \\ \times
e^{-i(k_y-k'_y)y}\, V(x,y)
\simeq V_Bk_xk'_x e^{-(k_y-k'_y)^2a^2},
\end{eqnarray}
For the probability of scattering from all such defects located in sublattices A and B at the
zigzag boundary we obtain
\begin{eqnarray}
\label{21}
W_z({\bf k,k'})=\frac{2\pi }{\hbar }\, N_iV_1^2k_x^2k'^2_x\,
e^{-2(k_y-k'_y)^2a^2}\,
\delta (\varepsilon _{\bf k}-\varepsilon _{\bf k'}), \hskip0.3cm
\end{eqnarray}
where we introduced the notation $N_iV_1^2\equiv N_{iA}V_A^2+N_{iB}V_B^2$,
$N_{iA}$ and $N_{iB}$ are the densities of impurities in sublattices
$A$ and $B$, respectively, and $N_i$ is the total density
of scatterers, $N_i=N_{iA}+N_{iB}$. One can assume $N_{iA}\simeq N_{iB}$.
We see that in this case (but not for the Berry-Mondragon boundary conditions!)
the scattering probability vanishes for the sliding electrons, $k_x \rightarrow 0$,
similar to the conventional metals.\cite{okulov79,falkovsky83} 

Using the same method as before we find for the zigzag boundary
\begin{eqnarray}
\label{22}
\mathcal{C}^>_z({\bf k})=\mathcal{C}^<_z({\bf k})=
\frac{eE\tau N_iV_1^2k_x^2}{2\pi \hbar ^2|v_x|\, (1-e^{-L/l_x})}
\left(-\frac{\partial f_0}{\partial \varepsilon }\right)
\nonumber \\
\times \int _{-k_F-k_y}^{k_F-k_y}
e^{-2q^2a^2}qdq\, [k_F^2-(k_y+q)^2]^{1/2},
\hskip0.2cm
\end{eqnarray}

\section{Conductance of the graphene nanoribbon}

The mean current density in the ribbon can be presented as $j=j_0+\delta j$, where
the average value is
\begin{eqnarray}
\label{23}
j=\frac{e}{2\pi ^2vL}\int _0^L dx
\int \varepsilon d\varepsilon \int \frac{k_ydk_y}{\hbar kk_x}
\left[ \delta f^>(k_y)+\delta f^<(k_y)\right] .\hskip0.3cm
\end{eqnarray}
It includes averaging over the ribbon width.
The term $j_0$, which does not depend on the edge scattering is
\begin{eqnarray}
\label{24}
j_0=\frac{e^2E}{\pi ^2v\hbar }
\int \varepsilon d\varepsilon \int \frac{k_ydk_y}{kk_x}\;
v_y\tau \left(-\frac{\partial f_0}{\partial \varepsilon }\right)
= \frac{e^2E\varepsilon _F \tau }{2\pi \hbar ^2},\hskip0.2cm
\end{eqnarray}
and $\delta j$ term is due to the edge ($\delta j<0$)
\begin{eqnarray}
\label{25}
\delta j=\frac{e}{\pi ^2v\hbar L}
\int \varepsilon d\varepsilon
\int \frac{k_ydk_y}{kk_x}\; l_x (1-e^{-L/l_x})\, \mathcal{C}^>(k_y) .\hskip0.3cm
\end{eqnarray}
As follows from (24), $\sigma _0=e^2\varepsilon _F\tau /2\pi \hbar ^2$ is the
conductance of infinite sample, $L\to \infty $.

In the case of Berry-Mondragon boundary conditions, substituting (17) in (25)
we obtain
\begin{eqnarray}
\label{26}
\delta j^{(BM)}
=j_0\, \frac{\tau N_iV_0^2k_F}{\pi ^2v\hbar L}
\int _{-1}^1\frac{\tilde{k}_y d\tilde{k}_y}{\sqrt{1-\tilde{k}_y^2}}
\nonumber \\ \times
\int _{-1-\tilde{k}_y}^{1-\tilde{k}_y}
\frac{e^{-2\xi \tilde{q}^2}\tilde{q}d\tilde{q}}{\sqrt{1-(\tilde{k}_y+\tilde{q})^2}},
\end{eqnarray}
where we denote
\begin{eqnarray}
\label{kssi}
\xi =a^2k_F^2, 
\end{eqnarray}
$\tilde{k}_y=k_y/k_F$, $\tilde{q}=q/k_F$.

In the case of zigzag boundary conditions we get
\begin{eqnarray}
\label{28}
\delta j^{(z)}
=j_0\, \frac{\tau N_iV_1^2k_F^5}{\pi ^2v\hbar L}
\int _{-1}^{1} \sqrt{1-\tilde{k}_y^2}\; \tilde{k}_yd\tilde{k}_y
\nonumber \\ \times
\int _{-1-\tilde{k}_y}^{1-\tilde{k}_y}
e^{-2\xi \tilde{q}^2} \sqrt{1-(\tilde{k}_y+\tilde{q})^2}\; \tilde{q}d\tilde{q}.
\end{eqnarray}
The dependence of conductivity on ~the ribbon width $L$ is shown in Figs.~1 and 2.
Here we used
\begin{eqnarray}
\label{29}
\sigma _{BM}=\sigma _0\left( 1+\frac{\gamma _{BM}\, \ell }{L}
\int _{-1}^1\frac{\tilde{k}_y d\tilde{k}_y}{\sqrt{1-\tilde{k}_y^2}}
\right. \nonumber \\ \times \left.
\int _{-1-\tilde{k}_y}^{1-\tilde{k}_y}
\frac{e^{-2\xi \tilde{k}_y^2}\, \tilde{q}d\tilde{q}}{\sqrt{1-(\tilde{k}_y+\tilde{q})^2}}\right) ,
\end{eqnarray}
\begin{eqnarray}
\label{30}
\sigma _z=\sigma _0\left( 1+\, \frac{\gamma _z\, \ell }{L}
\int _{-1}^1\frac{\tilde{k}_y d\tilde{k}_y}{\sqrt{1-\tilde{k}_y^2}}
\right. \nonumber \\ \times \left.
\int _{-1-\tilde{k}_y}^{1-\tilde{k}_y}
\frac{e^{-2\xi y^2}\, \tilde{q}d\tilde{q}}{\sqrt{1-(\tilde{k}_y+\tilde{q})^2}}\right) ,
\end{eqnarray}
with notations
\begin{eqnarray}
\label{31}
\gamma _{BM}=\frac{N_iV_0^2k_F}{\pi ^2v^2},
\hskip0.3cm
\gamma _z=\frac{N_iV_1^2k_F^5}{\pi ^2v^2},
\hskip0.3cm
\ell =\frac{v\tau }{\hbar },
\end{eqnarray}
and in Eqs.~(29) and (30) $\ell $ is the ``bulk'' mean path in graphene. Note that Eqs.~(29) and
(30) are valid only when the second term related to the edge scattering is a small
correction to the bulk conductivity, $|\Delta \sigma |\ll \sigma _0$.

In numerical calculations of Figs. 1 and 2 we choose the length unit $a_0=10^{-8}$~cm.
We also take $N_iV_0^2/\pi ^2v^2a_0^2=10^2$ and $N_iV_1^2/\pi ^2v^2a_0^6=10^8$.
It corresponds, e.g., to the following choice of parameters: $N_i=10^{-4}/a_0=10^4$~cm$^{-1}$, 
$V_0=0.1ta_0^2\simeq 3\times 10^{-17}$~eV$\cdot $cm$^2$, 
$V_1=10^2ta_0^4\simeq 3\times 10^{-30}$~eV$\cdot $cm$^4$, $v=10^{-8}$~eV$\cdot $cm.  
This choice provides fullfillment of the perturbation approximation condition
$|\Delta \sigma |\ll \sigma _0$. For the $\xi $ parameter we take $\xi =1$ (like for defects
in form of "steps" of the order of electron wave length).

\begin{figure}[ptb]
\hspace*{-1cm}
\includegraphics[width=0.9\linewidth]{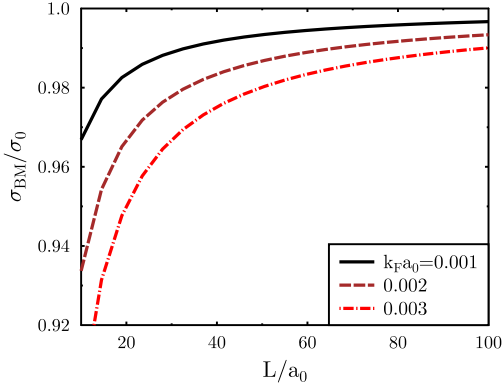}
\vspace*{-0.2cm}
\caption{(color online) Conductivity as a function of $L$ for different values of $k_F$
(Berry-Mondragon boundary conditions at the edges).
For numerical calculations we take $\gamma _{BM}=10^2(a_0/\ell )(k_Fa_0)$.}
\end{figure}

\begin{figure}[ptb]
\hspace*{-1cm}
\includegraphics[width=0.9\linewidth]{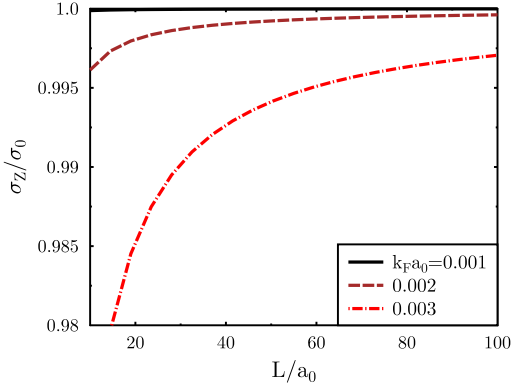}
\vspace*{-0.2cm}
\caption{(color online) Conductivity as a function of $L$ for different values of $k_F$
(zigzag boundary conditions at the edges). Here we take $\gamma _z=10^8(a_0/\ell )(k_Fa_0)^5$.}
\end{figure}

\section{Graphene nanoribbon in the ballistic regime}

Now we assume that there is no scatterers in the bulk.
It corresponds to the ballistic limit when the bulk mean free path $\ell $ is large
comparing to the ribbon width, $\ell \gg L$.
Then the kinetic equation for the distribution function in the bulk
includes only the scattering from the edges
\begin{eqnarray}
\label{32}
eEv_y\, \frac{\partial f_0}{\partial \varepsilon }
=\sum _{\bf k'} W_{\bf kk'}
(f^{<,>}_{\bf k'}-f^{>,<}_{\bf k})
\end{eqnarray}
where $W({\bf k,k'})$ is the probability of edge scattering.

Using Eqs.~(32) we can decouple them as an equation for $f^>_{\bf k}$ and 
another equation for $f^<_{\bf k}$, from which follows that in the ballistic regime 
$f^>_{\bf k}=f^<_{\bf k}$. Thus, in this regime we drop out the
"forward" and "backward" indices.
As before, we can find the solutions of these equations by using the boundary condition
for the wave function of different type.

\subsection{Solution for the Berry-Mondragon boundary}

In the case of Berry-Mondragon boundary conditions, Eq.~(32) with 
$f^>_{\bf k}=f^<_{\bf k}=f_{\bf k}$ can be written as
\begin{eqnarray}
\label{33}
eEv_y\, \frac{\partial f_0}{\partial \varepsilon }
=\frac{2\pi N_iV_0^2}{\hbar L}
\int \frac{d^2{\bf k'}}{(2\pi )^2}\;
e^{-2(k_y-k'_y)^2a^2}\delta (\varepsilon _{\bf k}-\varepsilon _{\bf k'})
\nonumber \\ \times \,
(f_{\bf k'}-f_{\bf k}).\hskip0.5cm
\end{eqnarray}
The solution of Eq. (33) has the following form
\begin{eqnarray}
\label{34}
f_{\bf k}=eEv_y\left( -\frac{\partial f_0}{\partial \varepsilon }\right) \tau _{BM}(k_y),
\end{eqnarray}
where $\tau _{BM}(k_y)$ is the relaxation time depending on the angle,
under which electrons are incoming to the edge, and $\tau _{BM}(k_y)=\tau _{BM}(-k_y)$.
Substituting Eq. (34) into Eq. (33) we obtain an equation for the function $\tau _{BM}(k_y)$.

If the parameter $\xi\equiv a^2k_F^2\ll 1$ (which is a realistic case, if $a$ is of the order 
of several interatomic distances), 
this equation can be solved analytically.
In this case the dependence of $\tau _{BM}$ on $k_y$ turns out to be weak.
Therefore, the equation for $\tau _{BM}$ reduces to
\begin{eqnarray}
\label{35}
\frac1{\tau _{BM}}
=-\frac{N_iV_0^2k_F}{2\pi \hbar Lv}\, \frac1{\tilde{k}_y}
\int _{-1-\tilde{k}_y}^{1-\tilde{k}_y}
\frac{e^{-2\xi \tilde{q}^2}
\tilde{q}d\tilde{q}}{\sqrt{1-(\tilde{k}_y+\tilde{q})^2}} .\hskip0.3cm
\end{eqnarray}

\begin{figure}[ptb]
\hspace*{-1cm}
\includegraphics[width=1.1\linewidth]{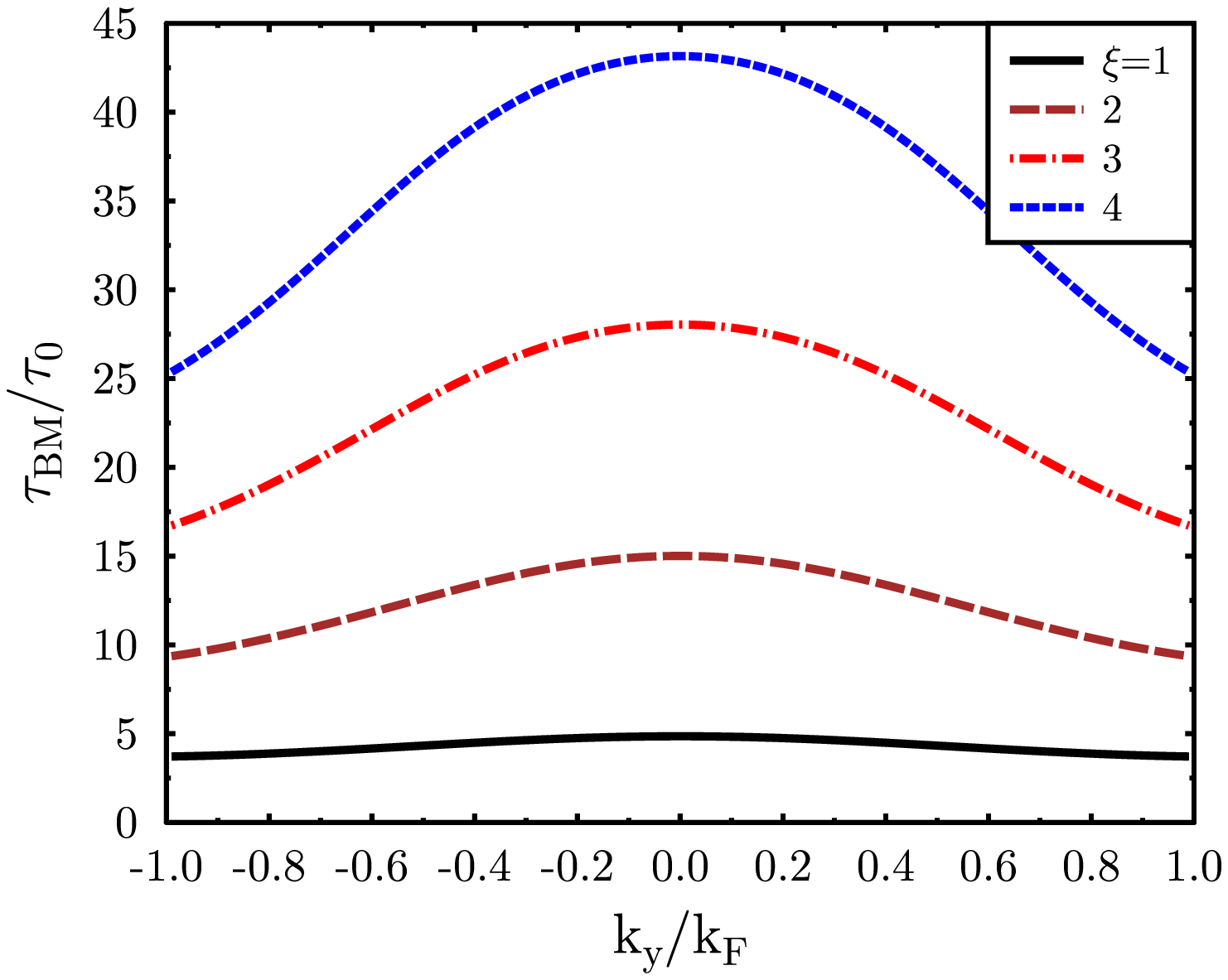}
\vspace*{-0.2cm}
\caption{(color online) Relaxation time as a function of $k_y/k_F$ for different values of the parameter
$\xi =a^2k_F^2$ (Berry-Mondragon boundary conditions at the edges).}
\end{figure}

For arbitrary (not necessarily small) value of the parameter $\xi $
we can present the equation for $\tau _{BM}(k_y)$ in the following form
\begin{eqnarray}
\label{36}
\int _{-1}^{1}d\tilde{k}'_y\, e^{-2\xi (\tilde{k}_y-\tilde{k}'_y)^2}\,
\frac{\tilde{k}_y\tilde{\tau }_{BM}(\tilde{k}_y)-\tilde{k}'_y\tilde{\tau }_{BM}(\tilde{k}'_y)}
{\tilde{k}_y \tilde{k}'_x}=1,
\end{eqnarray}
where $\tilde{\tau }_{BM}=\tau _{BM}/\tau _0$, and $\tau _0^{-1}=N_iV_0^2k_F/2\pi \hbar Lv$.
Thus, we find
\begin{eqnarray}
\label{37}
\tilde{\tau }_{BM}(\tilde{k}_y)=\frac{1+\int _{-1}^1 d\tilde{k}'_y\, e^{-2\xi (\tilde{k}_y-\tilde{k}'_y)^2}
\frac{\tilde{k}_y\tilde{\tau }_{BM}(\tilde{k}_y)}{\tilde{k}_y\tilde{k}'_x}}
{\int _{-1}^1 d\tilde{k}'_y\, e^{-2\xi (\tilde{k}_y-\tilde{k}'_y)^2}\frac1{\tilde{k}'_x}} .
\end{eqnarray}
Solving Eq.~(37) self-consistently by iterations, we find the dependence
$\tilde{\tau }_{BM}(\tilde{k}_y)$.
This solution is presented in Fig.~3. It shows that the transport relaxation time of
electrons incoming under small angles ($|k_y|/k_F\sim 1$) is smaller that those incoming under
large angles, and this effect is more significant for large $\xi $ (i.e., when the electron
wavelength $\lambda $ is small with respect to the characteristic dimension of imperfections,
$\lambda \ll a$). In other words, in the case of Berry-Mondragon boundary, sliding electrons are
scattered from edges more effectively. This is because the electron wave function is not
zero at the edge.

\subsection{Solution for the zigzag edge}

In the case of zigzag edge, using Eq. (22) and calculating the electron relaxation time like
before, for $\xi \ll 1$ we find the solution in the following analytical form
\begin{eqnarray}
\label{38}
\frac1{\tau _z}
=-\frac{N_iV_1^2k_F^5}{2\pi \hbar Lv}\, \frac{\tilde{k}_x^2}{\tilde{k}_y}
\int _{-1-\tilde{k}_y}^{1-\tilde{k}_y}e^{-2\xi \tilde{q}^2}\sqrt{1-(\tilde{k}_y+\tilde{q})^2}\;
\tilde{q}d\tilde{q}.\hskip0.2cm
\end{eqnarray}
For arbitrary $\xi $ we find the following equation for $\tau _z(k_y)$
\begin{eqnarray}
\label{39}
\tilde{\tau }_z(\tilde{k}_y)=\frac{1+\frac{\tilde{k}_x^2}{\tilde{k}_y}\int _{-1}^1 d\tilde{k}'_y\,
e^{-2\xi (\tilde{k}_y-\tilde{k}'_y)^2}
\tilde{k}'_y\tilde{k}'_x\, \tilde{\tau _z}(\tilde{k}'_y)}
{\tilde{k}_x^2\int _{-1}^1 d\tilde{k}'_y\, e^{-2\xi (\tilde{k}_y-\tilde{k}'_y)^2}\tilde{k}'_x} ,
\end{eqnarray}
where we denote $\tilde{\tau }_z=\tau _z/\tau _1$, and $\tau _1^{-1}=N_iV_1^2k_F^5/2\pi \hbar Lv$.

\begin{figure}[ptb]
\hspace*{-1cm}
\includegraphics[width=1.1\linewidth]{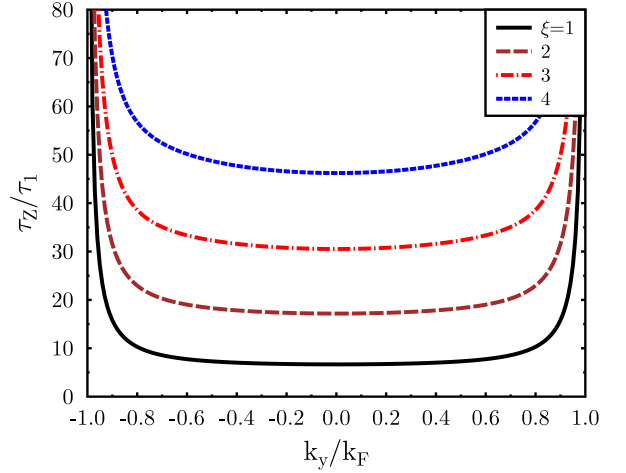}
\vspace*{-0.2cm}
\caption{(color online) Relaxation time as a function of $k_y/k_F$ for different values of the parameter $\xi $
(zigzag boundary conditions at the edges).}
\end{figure}

Solving Eq.~(39) by iteration we find the dependence presented in Fig.~4.
As we see, in the case of zigzag boundary, sliding electrons with $|k_y|/k_F\sim 1$ do not scatter
from the edge at any value of the parameter $\xi $.
It means that the approximation of constant $\tau _z$ and solution (38) are not valid in
close vicinity to $|\tilde{k}_y|=1$ even for small $\xi $.

\subsection{Conductance in ballistic regime}

The conductance of carbon nanoribbon can be found now in the case of
Berry-Mondragon boundary and for the zigzag edges.
We can find, respectively,
\begin{eqnarray}
\label{40}
\sigma _{BM}
=\frac{4e^2v^2L}{N_iV_0^2}
\int _0^1 \frac{\tilde{k}_y^2\, \tilde{\tau }_{BM}(\tilde{k}_y)\, d\tilde{k}_y}
{\sqrt{1-\tilde{k}_y^2}}
\end{eqnarray}
where $\tilde{\tau }_{BM}=\tau _{BM}/\tau _0$ and
\begin{eqnarray}
\label{41}
\frac1{\tilde{\tau }_{BM}(\tilde{k}_y)}=-\frac{1}{\tilde{k}_y}\int _{-1-\tilde{k}_y}^{1-\tilde{k}_y}
\frac{e^{-2\xi \tilde{q}^2}\tilde{q}d\tilde{q}}{\sqrt{1-(\tilde{k}_y+\tilde{q})^2}},
\end{eqnarray}
and
\begin{eqnarray}
\label{42}
\sigma _{z}
=\frac{4e^2v^2L}{N_iV_1^2k_F^4}
\int _0^1 \frac{\tilde{k}_y^2\, \tilde{\tau }_z(\tilde{k}_y)\, d\tilde{k}_y}{\sqrt{1-\tilde{k}_y^2}},
\end{eqnarray}
where $\tilde{\tau }_z=\tau _z/\tau _1$ and
\begin{eqnarray}
\label{43}
\frac1{\tilde{\tau }_z(\tilde{k}_y)}=-\frac{\tilde{k}_x}{\tilde{k}_y}
\int _{-1-\tilde{k}_y}^{1-\tilde{k}_y} e^{-2\xi \tilde{q}^2}\sqrt{1-(\tilde{k}_y+\tilde{q})^2}\;
\tilde{q}d\tilde{q} .
\hskip0.3cm
\end{eqnarray}

\section{Graphene nanoribbon with curved edges}

Now we consider the case of curved edges of the ribbon.
Let the left edge is now at $x=-L/2+s_1(y)$ and the right
edge $x=L/2+s_2(y)$, where $s_1(y)$ and $s_2(y)$ are some arbitrary
functions characterizing disorder of the ribbon edge.
We assume $|s_{1,2}(y)|\ll L$, and disorder properties of
$s_1(y)$ and $s_2(y)$ completely uncorrelated.

It is convenient to introduce new coordinates ($x',y'$) using
conformal transformation
\begin{eqnarray}
\label{44}
x=x'+s(y)+\alpha (y)\, x',
\hskip0.5cm
y=y',
\end{eqnarray}
where $s(y)=(s_1+s_2)/2$ and $\alpha (y)=(s_2-s_1)/L$.
As follows from this definition, each point at the left
edge with $x=-L/2+s_1(y)$ corresponds to $x'=-L/2$,
and each point at the right edge with $x=L/2+s_2(y)$ corresponds
to $x'=L/2$. In other words, in new ($x',y'$) coordinates the edges
of ribbon are strait lines.

In correspondance with (44) we find ($|\alpha |\ll 1$)
\begin{eqnarray}
\label{45}
x'=\frac{x-s}{1+\alpha }\simeq (x-s)(1-\alpha ),
\hskip0.5cm
y'=y.
\end{eqnarray}
The transformation of derivatives is
\begin{eqnarray}
\label{46}
\partial _x\simeq (1-\alpha )\, \partial _{x'}\,
\hskip0.3cm
\partial _y\simeq -s'\, \partial _{x'}+\partial _{y'}\, ,
\end{eqnarray}
where $s'\equiv ds/dy$ and $\alpha '\equiv d\alpha /dy$.

The Dirac Hamiltonian in new coordinates is
$H=H_0+H^{(C)}_{int}$, where $H_0=-iv\sigma _i\partial _i$ and
\begin{eqnarray}
\label{47}
H^{(C)}_{int}=iv\alpha \sigma _x\, \partial _x
+ivs'\sigma _y\, \partial _x
\end{eqnarray}
is the perturbation related to the curved edges.
As follows from (47) the above-mentioned coordinate transformation generates
the following gauge field
\begin{eqnarray}
\label{48}
{\bf A}=i\left( \alpha \partial _x,\; s' \partial _x\right) .
\end{eqnarray}
Perturbation (47) leads to nonzero matrix elements of transitions
between eigenstates ($k_x,k_y$) and ($k_x,k'_y$) of the Hamiltonian $H_0$.
Due to elasticity of scattering we should take into account only backscattering
transitions with $k_y\to -k_y$, which contribute to the transport properties of
graphene nanoribbon.

Matrix elements of transition ${\bf k} \to {\bf k+q}$ with ${\bf q}=(0,\, q)$, are
(here we use Berry-Mondragon condition for the wavefunction)
\begin{eqnarray}
\label{49}
\left< {\bf k}|H^{(C)}_{int}|{\bf k+q}\right>
=-\frac{\alpha _qvk_xk_-}{k\mathcal{L}}\, ,
\end{eqnarray}
where $\alpha _q=\int \alpha (y)\, e^{-iqy}dy$ and $\mathcal{L}$ is the ribbon length.

Now the right-hand part of kinetic equation is
\begin{eqnarray}
\label{50}
{\rm St}\, f_{\bf k}
=-\frac{2\pi }{\hbar }\sum _{\bf k'}
\left| \left< {\bf k}|H^{(C)}_{int}|{\bf k'}\right> \right| ^2
\delta (\varepsilon _{\bf k}-\varepsilon _{\bf k'})
(f_{\bf k}-f_{\bf k'})
\nonumber \\
=-\frac{|\alpha _{2k_y}|^2 vk_x^2k}{\hbar k_y\mathcal{L}}\,
(f_{\bf k}-f_{\bf k+q_0}).\hskip0.3cm
\end{eqnarray}
Here $f_{\bf k}=f(k_x,k_y)$ and $f_{\bf k+q_0}=f(k_x,-k_y)$.

Averaging over realizations of $\alpha (y)$ gives us
\begin{eqnarray}
\label{51}
\overline{|\alpha _q|^2}=\int dy\, dy'\, e^{iq(y-y')}\, \overline{\alpha (y)\, \alpha (y')}
=\mathcal{L}\, C_q\, ,
\end{eqnarray}
where we denote $C_q=\int dy\, e^{iqy}\, \overline{\alpha (y)\, \alpha (0)}$.
In the following we can assume $C_q=\left< \alpha ^2\right> a_\alpha \exp{(-a_\alpha ^2q^2)}$,
where $a_\alpha $ is the characteristic length of fluctuations.

Then after averaging we obtain
\begin{eqnarray}
\label{52}
{\rm St}\, f_{\bf k}
=-\frac{C_{2k_y} vk_x^2k}{\hbar k_y}\,
(f_{\bf k}-f_{\bf k+q_0}),\hskip0.3cm
\end{eqnarray}
and the kinetic equation acquires the form
\begin{eqnarray}
\label{53}
eEv_y\, \frac{\partial f_0}{\partial \varepsilon }
=-\frac{C_{2k_y}vk_x^2k}{\hbar k_y}\,
(f_{\bf k}-f_{\bf k+q_0}).\hskip0.3cm
\end{eqnarray}
Hence, one can identify the relaxation time as
$\tau _{\bf k}^{-1}=C_{2k_y}vk_x^2k/\hbar k_y$.

Electric current along the ribbon is
\begin{eqnarray}
\label{54}
j=\frac{2ev}{\hbar }\int \frac{d^2{\bf k}}{(2\pi )^2}\frac{k_y}{k}\, f_{\bf k}
\simeq \frac{e}{2\pi ^2\hbar v}
\int _0^\infty \frac{k_ydk_y}{kk_x}
\nonumber \\ \times
\int \varepsilon d\varepsilon \, (f_{\bf k}-f_{\bf k+q_0}).
\end{eqnarray}
Using (53) and (54) we find the conductance determined by the curved edges
\begin{eqnarray}
\label{55}
\sigma _C=\frac{e\varepsilon _F}{2\pi ^2v\hbar }
\int _0^{k_F} \frac{k_ydk_y}{kk_x}\,
\frac{ev_y\hbar k_y}{C_{2k_y}vk_x^2k}
=\frac{e^2\varepsilon _FL}{4\pi ^2v\hbar k_FC_{2k_F}}\, ,\hskip0.2cm
\end{eqnarray}
where we have to cut integral at small $k_x$ by $k_{min}\simeq 1/L$.

Combining $\sigma _C$ with the conductivity of graphene
without curved edges $\sigma _0$ and assuming $\sigma _C\gg \sigma _0$
we obtain
\begin{eqnarray}
\label{56}
\sigma \simeq \sigma _0\, ( 1-\sigma _0/\sigma _C).
\end{eqnarray}
Then using Eq.~(55) we get
\begin{eqnarray}
\label{57}
\sigma \simeq \sigma _0\left( 1-\frac{\pi \tau vk_F}{\hbar L}
\left< \alpha ^2\right> a_\alpha e^{-4a_\alpha ^2k_F^2}\right) .
\end{eqnarray}
Formula (57) presents the correction to conductance related to the curved
edges if $\sigma _C/\sigma _0\ll 1$. In the opposite case of ballistic
ribbon, the conductance is presented by Eq.~(55).

\section{Intervalley transitions due to the scattering from the random gauge potential}

Our approach can be generalized to take into account possible intervalley
transitions. For this purpose we can use full Hamiltonian of graphene in tight-binding
approximation, which describes the states in the whole Brillouin zone \cite{katsnelson}
\begin{eqnarray}
\label{58}
H_0=\left( \begin{array}{cc} 0 & t\, \xi ({\bf k}) \\ t\, \xi ^*({\bf k}) & 0 \end{array}\right) ,
\end{eqnarray}
where
\begin{eqnarray}
\label{59}
\xi ({\bf k})=2\cos \left( \frac{k_ya\sqrt{3}}2 \right) e^{ik_xa/2}+e^{-ik_xa} ,
\end{eqnarray}
$t$ is the hopping energy and $a$ is the lattice constant.
The Dirac points $\mathcal{K}$ and $\mathcal{K'}$ correspond to two nonequivalent points of the Brillouin zone,
at which $\xi ({\bf k})=0$
\begin{eqnarray}
\label{60}
{\bf K}=\left( \frac{2\pi }{3a},\; -\frac{2\pi }{3\sqrt{3}a}\right) ,
\hskip0.5cm
{\bf K'}=\left( \frac{2\pi }{3a},\; \frac{2\pi }{3\sqrt{3}a}\right) .
\end{eqnarray}
By using the coordinate transformation (44) we obtain the perturbation
\begin{eqnarray}
\label{61}
H_{int}=\left( \begin{array}{cc} 0 & tA_i \xi _i \\
tA_i^* \xi ^*_i & 0 \end{array} \right) ,
\end{eqnarray}
where we denoted
\begin{eqnarray}
\label{62}
&&\xi _x\equiv \frac{\partial \xi }{\partial k_x}
=ia\left[ \cos \left( \frac{k_ya\sqrt{3}}2 \right) e^{ik_xa/2}-e^{-ik_xa}\right] ,
\nonumber \\
&&\xi _y\equiv \frac{\partial \xi }{\partial k_y}
=-a\sqrt{3}\, \sin \left( \frac{k_ya\sqrt{3}}2 \right) e^{ik_xa/2} ,
\end{eqnarray}
the vector ${\bf k}$ should be understood as the momentum operator, and ${\bf A}$ is
defined by Eq.~(48).

We need to calculate interband matrix elements of the perturbation (61) with the wave functions
of electrons in valleys $\mathcal{K}$ and $\mathcal{K'}$
\begin{eqnarray}
\label{63}
&&\left| \tilde{\bf k},\mathcal{K}\right> =\frac{e^{i({\bf K}+\tilde{\bf k})\cdot {\bf r}}}{\sqrt{2\Omega }}
\left( \begin{array}{c} 1 \\ \tilde{k}_+/\tilde{k}\end{array}\right) ,
\\
&&\left| \tilde{\bf k'},\mathcal{K'}\right> =\frac{e^{i({\bf K'}+\tilde{\bf k'})\cdot {\bf r}}}{\sqrt{2\Omega }}
\left( \begin{array}{c} 1 \\ \tilde{k}'_-/\tilde{k}'\end{array}\right) ,
\end{eqnarray}
where $\tilde{\bf k}$ and $\tilde{\bf k'}$ are the electron momenta measured from
the Dirac points $\mathcal{K}$ and $\mathcal{K'}$, respectively.

The interband transition is nonzero if it conserves the $x$-component of moment,
$K_x=K'_x$, $\tilde{k}_x=\tilde{k}'_x$, and corresponds to the transfer with $K_y=K'_y\pm Q$, where
$Q=4\pi /3\sqrt{3}a$. As before, due to the elasticity of scattering, we can consider
only the matrix elements of intervalley transitions between $\tilde{\bf k}$ and
$\tilde{\bf k}'=\tilde{\bf k}+{\bf q}$ with ${\bf q}=(0,q)$ (intervalley backscattering),
so that both $\tilde{\bf k}$ and $\tilde{\bf k}+{\bf q}$ are at the same energy surface.

Using Eqs.~(61)-(64) with gauge filedl (48) and assuming $\tilde{k},q\ll Q$ we obtain
\begin{eqnarray}
\label{65}
\left< \tilde{\bf k},\mathcal{K}\right| H_{int}\left| \tilde{\bf k}+{\bf q},\mathcal{K'}\right> \hskip5cm
\nonumber \\
\simeq \frac{t}{2}\left(
\left\{ -i\alpha _QK_xa\left[ \cos \left( \frac{K'_ya\sqrt{3}}2\right)
e^{iK_xa/2}-e^{-iK_xa}\right]
\right. \right. \nonumber \\ \left. \left.
-s'_QK_xa\sqrt{3}\sin \left( \frac{K'_ya\sqrt{3}}2\right) e^{iK_xa/2}
\right\} \frac{k'_-}{k'}
\right. \nonumber \\ \left.
+\left\{ i\alpha _QK_xa\left[ \cos \left( \frac{K'_ya\sqrt{3}}2\right)
e^{-iK_xa/2}-e^{iK_xa}\right]
\right. \right. \nonumber \\ \left. \left.
-s'_QK_xa\sqrt{3}\sin \left( \frac{K'_ya\sqrt{3}}2\right) e^{-iK_xa/2}
\right\} \frac{k_-}{k}\right)
\hskip0.3cm
\nonumber \\
=\frac{\pi t}{k}\left[ \frac{\alpha _Q}2 \left( \frac{5k_x}{\sqrt{3}} -k_y \right)
-s'_Q(k_x-k_y)\right] ,\hskip0.3cm
\end{eqnarray}
where $\alpha _Q=\int \alpha (y)\, e^{-iQy}dy$, $s'_Q=\int s'(y)\, e^{-iQy}dy$,
and $Q=K_y-K'_y=-4\pi /3\sqrt{3}a$.  

Then using the same method as in Sec.~VI, we find the conductance limited by 
intervalley scattering from the fluctuating gauge potential
\begin{eqnarray}
\label{66}
\sigma _{iv}
=\frac{e^2\varepsilon _Fv}{4\pi ^4t^2\hbar }
\int _0^{1} 
\frac{\tilde{k}_y^3d\tilde{k}_y}{\tilde{k}_x
\left[ \frac{C_Q}4 \left( \frac{5\tilde{k}_x}{\sqrt{3}} -\tilde{k}_y\right) ^2
+R_Q\, (\tilde{k}_x-\tilde{k}_y)^2\right] },\hskip0.3cm
\end{eqnarray} 
where $C_Q=\overline{|\alpha _Q|^2}/\mathcal{L}$ and
$R_Q=\overline{|s'_Q|^2}/\mathcal{L}$ are the correlators of randomly 
fluctuating fields $\alpha (y)$ and  $s'(y)$.

Correspondingly, the intervalley relaxation time related to this mechanism is
\begin{eqnarray}
\label{67}
\tau _{iv}=\frac{v\hbar }{\pi ^2t^2} 
\int _0^{1} 
\frac{\tilde{k}_y^3d\tilde{k}_y}{\tilde{k}_x
\left[ \frac{C_Q}4 \left( \frac{5\tilde{k}_x}{\sqrt{3}} -\tilde{k}_y\right) ^2
+R_Q\, (\tilde{k}_x-\tilde{k}_y)^2\right] }.\hskip0.3cm
\end{eqnarray} 
Note that this type of interband transition mechanism can be realized for sufficiently
sharp-curved edges because it is associated with the large transfered momentum $Q$.

\section{Intervalley transitions due to the wavefunction boundary condition at the edge}

In the case of a reconstructed zigzag edge, the most energetically stable is
zz(57) or ``reczag'' reconstruction.\cite{koskinen08} In this case, corresponding boundary
conditions at the edge are equivalent to additional intervalley-inducing
term in the Dirac Hamiltonian \cite{katsnelson,ostaay11}
\begin{eqnarray}
\label{68}
H_{iv}=vM'\, \delta [x-s(y)] ,
\end{eqnarray}
where we assume the edge at $x=s(y)$.
Matrix $M'$ in (68) is Hermitian and acts in spaces of valleys and sublattices.
It leads to the boundary condition for the wave function at the edge
\begin{eqnarray}
\label{69}
\psi =M\psi .
\end{eqnarray}
Matrices $M$ and $M'$ are connected through
\begin{eqnarray}
\label{70}
M=i\tau _0 \sigma _xM' .
\end{eqnarray}
For the reczag reconstruction the matrix $M$ is
\begin{eqnarray}
\label{71}
M=(\bnu\cdot \btau )\, ({\bf n}\cdot \bsig ) ,
\end{eqnarray}
where $\bnu $, $\bf n$ are some unit vectors, ${\bf n}\bot {\bf n}_B$, and ${\bf n}_B$ is the
unit vector normal to the boundary. Pauli matrices $\btau $ and $\bsig $ refer to
the valley and sublattice spaces, respectively.

If the edge is flat, $s(y)=0$, then due to the chiral symmetry we should
take for the reczag reconstruction $\bnu || \hat{z}$ and ${\bf n}$ in the
$y$-$z$ plane. We obtain
\begin{eqnarray}
\label{72}
M=-\tau _z\, (\sigma _z\cos \theta +\sigma _y\sin \theta ),
\end{eqnarray}
and the angle $\theta =0.150$.\cite{ostaay11}
Corresponding Hamiltonian does not couple different valleys.

In the absence of chiral symmetry one can use general form (71) of $M$.
Assuming deviation from the flat edge small, we can consider
curvature-unduced "interaction" term in the matrix $M$
\begin{eqnarray}
\label{73}
M_{int}=s'(y)\, (\beta _1\tau _x+\beta _2\tau _y)\,
(\sigma _z\cos \theta +\sigma _y\sin \theta ) ,
\end{eqnarray}
where $\beta _1$, $\beta _2$ are come coefficients determined by the specific
reconstruction type at the edge, and we assume
$s'(y)\, \beta _{1,2}\ll 1$. These terms induce intervalley transitions.
Correspondingly, we obtain from (68),(70) and (73)
\begin{eqnarray}
\label{74}
H_{int}=vs'(y)\, (\beta _1\tau _x+\beta _2\tau _y)\,
(-\sigma _y\cos \theta +\sigma _z\sin \theta )\,
\nonumber \\ \times
\delta [x-s(y)] .
\end{eqnarray}
As we see, this perturbation couples different valleys leading
to intervalley transitions. In other words, it means edge-induced
valley relaxation.

The conductance limited by intervalley transitions resulting from the scattering
of the reconstructed edge can be calculated as in Sec.~VI. We find 
\begin{eqnarray}
\label{75}
\sigma _{rec}
=\frac{e^2\varepsilon _F}{4\pi ^2v\hbar k_F^2R_{2k_F}(\beta _1^2+\beta _2^2)\sin ^2\theta }.
\end{eqnarray}
Correspondingly, we can find the intervalley relaxation time
\begin{eqnarray}
\label{76}
\tau _{rec}=\frac{\hbar }{vk_F^2R_{2k_F}(\beta _1^2+\beta _2^2)\sin ^2\theta }.
\end{eqnarray}
It should be noted that both Eqs.~(67) and (76) describe the 'intervalley transport' relaxation time
as they are associated with the backscattering, $k_y\to -k_y$ of electrons.

\section{Ballistic disc}

Now we consider edge-induced relaxation of the electron momentum in
a ballistic disc.
In the case of a disc of radius $R$, instead of cartesian $x,y$
it is more convenient to use polar coordinates $r,\phi $. Then
the Schr\"odinger equation for $r<R$ acquires the form
\begin{eqnarray}
\label{77}
&&\varepsilon \varphi
+ive^{-i\phi }\left( \frac{\partial }{\partial r}
-\frac{i}{r}\frac{\partial }{\partial \phi }\right) \chi =0,
\\
&&ive^{i\phi }\left( \frac{\partial }{\partial r}
+\frac{i}{r}\frac{\partial }{\partial \phi }\right) \varphi
+\varepsilon \chi =0.
\end{eqnarray}
We make the substitutions $\varphi (r,\phi )=e^{im\phi }\varphi _m(r) $ and
$\chi _{m+1}(r,\phi )=e^{i(m+1)\phi }\chi _{m+1}(r)$.
The solutions for $\varphi _m$ and $\chi _m$ are the Bessel functions $J_m(z)$ and $Y_m(z)$
with argument $z=r\varepsilon /v$.
They have asymptotics for large $z\gg 1$
\begin{eqnarray}
\label{79}
J_m(z)\simeq \sqrt{\frac{2}{\pi z}}\;
\cos \left( z-\frac{m\pi }2-\frac{\pi }4\right) ,
\\
Y_m(z)\simeq \sqrt{\frac{2}{\pi z}}\;
\cos \left( z-\frac{m\pi }2+\frac{\pi }4\right) .
\end{eqnarray}
We can use these asymptotics as we are interested in behavior of
the wave functions near the disc edge, i.e., for
$r\sim R\gg k_F^{-1}=v/\varepsilon _F$.

Thus we find for the spinor components of the eigenfunctions
\begin{eqnarray}
\label{81}
&&\varphi _{m}^{\pm }(r.\phi )
\simeq e^{im\phi }\sqrt{\frac{2v}{\pi r\varepsilon }}\;
\cos \left( \frac{r\varepsilon }{v}-\frac{m\pi }2 \pm \frac{\pi }4\right) ,
\\
&&\chi _{m+1}^{\pm }(r.\phi )
\simeq ie^{i(m+1)\phi }\sqrt{\frac{2v}{\pi r\varepsilon }}\;
\sin \left( \frac{r\varepsilon }{v}-\frac{m\pi }2\pm \frac{\pi }4\right) .\hskip0.6cm
\end{eqnarray}
Correspondingly, the eigenfunctions at $r\leq R$ ($r\approx R$) are
\begin{eqnarray}
\label{83}
\psi _m^{\pm }(r,\phi )
\simeq e^{im\phi }\sqrt{\frac{2v}{\pi r\varepsilon }}\;
\left( \begin{array}{c}
\cos \left( \frac{r\varepsilon }{v}-\frac{m\pi }2 \pm \frac{\pi }4\right) \\
ie^{i\phi }\sin \left( \frac{r\varepsilon }{v}-\frac{m\pi }2\pm \frac{\pi }4\right)
\end{array} \right) .\hskip0.2cm
\end{eqnarray}
Now we use the Berry-Mondragon boundary conditions for the wave functions at the disc edge.
\footnote{Obviously, we cannot consider zigzag boundary conditions for the whole perimeter of the disc.}

In the case of Berry-Mondragon boundary conditions, the equations for $r>R$ (in vacuum), assuming
$\varepsilon \ll \Delta _0$
\begin{eqnarray}
\label{84}
&&-\Delta _0\varphi _m
+iv\left( \frac{d}{dr}
+\frac{m+1}{r}\right) \chi _{m+1}=0,
\\
&&iv\left( \frac{d}{dr}
-\frac{m}{r}\right) \varphi _m .
+\Delta _0\chi _{m+1}=0.
\end{eqnarray}
It gives us as the solution, decreasing with $z=r\Delta _0/v$, the  modified Bessel functions
$K_m(z)$ with asymptotics for $z\gg 1$
\begin{eqnarray}
\label{86}
K_m(z)\simeq \sqrt{\frac{2}{\pi z}}\; e^{-z}.
\end{eqnarray}
Correspondingly we take the wavefunction at $r>R$
\begin{eqnarray}
\label{87}
\psi _m(r,\phi )
=B\sqrt{\frac{2v}{\pi r\Delta _0}}\;
e^{-(r-R)\Delta _0/v+im\phi }
\left( \begin{array}{c} 1 \\ ie^{i\phi } \end{array} \right) ,
\end{eqnarray}
where $B$ is a constant.
Using (83) and (87) and matching these spinor components at $r=R$ we obtain
\begin{eqnarray}
\label{88}
\frac{A_+}{\sqrt{\varepsilon }}\, \cos \left( \frac{R\varepsilon }{v}-\frac{m\pi }2
+\frac{\pi }4\right) \hskip2.5cm
\nonumber \\
+\frac{A_-}{\sqrt{\varepsilon }}\, \cos \left( \frac{R\varepsilon }{v}-\frac{m\pi }2 -\frac{\pi }4\right)
=\frac{B}{\sqrt{\Delta _0}}\, ,
\\
\frac{A_+}{\sqrt{\varepsilon }}\, \sin \left( \frac{R\varepsilon }{v}-\frac{m\pi }2 +\frac{\pi }4\right) \hskip2.5cm
\nonumber \\
+\frac{A_-}{\sqrt{\varepsilon }}\, \sin \left( \frac{R\varepsilon }{v}-\frac{m\pi }2 -\frac{\pi }4\right)
=\frac{B}{\sqrt{\Delta _0}}\, .
\end{eqnarray}
This leads to simple equation relating $A_+$ and $A_-$ coefficients: $A_+=-A_-$.

Finally, the wavefunction at $r<R$ obeying the Berry-Mondragon boundary condition is
\begin{eqnarray}
\label{90}
\psi _{km}(r,\phi )
=\frac{A_k e^{im\phi }}{\sqrt{k}}\;
\left( \begin{array}{c}
-\sin \left( kr-m\pi /2\right) \\
ie^{i\phi }\cos \left( kr-m\pi /2\right) 
\end{array} \right) ,
\end{eqnarray}
where $k=\varepsilon /v$ and $A_k$ is the normalization constant, $A_k\simeq k^{1/2}/R$.

Matrix elements of impurity potential, located at the edge of disc
in the sublattice A
\begin{eqnarray}
\label{91}
\left< km|V^{(A)}(r,\phi )|k'm'\right>
\simeq \frac{V_0}{R}
e^{-(m-m')^2a^2/R^2}
\nonumber \\ \times
\sin \left( kR-\frac{m\pi }2\right)
\sin \left( k'R-\frac{m'\pi }2\right) ,
\end{eqnarray}
Analogously, for the impurity localized in the sublattice B at the edge we get
\begin{eqnarray}
\label{92}
\left< km|V^{(B)}(r,\phi )|k'm'\right>
\simeq \frac{V_0}{R}
e^{-(m-m')^2a^2/R^2}
\nonumber \\ \times
\cos \left( kR-\frac{m\pi }2\right)
\cos \left( k'R-\frac{m'\pi }2\right) ,
\end{eqnarray}
The relaxation time can be evaluated from
\begin{eqnarray}
\label{93}
\frac{\hbar }{\tau _{k}}\simeq N_i \sum _{m's}\int dk'\,
\left| \left< km|V^{(s)}(r,\phi )|k'm'\right> \right| ^2
\nonumber \\ \times
\delta (\varepsilon _{k}-\varepsilon _{k'}),
\end{eqnarray}
where $N_i$ is the linear density of impurities at the edge of disc and $s=A,B$.
Using (91),(92) and assuming $k,k'\gg 1/R$ and one can finaly obtain
\begin{eqnarray}
\label{94}
\frac1{\tau _{k}}\simeq \frac{N_iV_0^2}{2\hbar R^2v}\sum _{m'}
e^{-2(m-m')^2a^2/R^2}
\simeq \frac{\sqrt{\pi }N_iV_0^2}{\sqrt{2}\, \hbar Rav}.
\end{eqnarray}

\section{Discussion of results}

The effect of surface scattering on the conductivity of thin films and wires has been considered first
\cite{lovell36,fuchs38,dingle50} by using
a constant specular factor $p$, which characterizes scattering properties of the surface,
so that the value of $p=0$ corresponds to specular scattering and $p=1$ to the
diffusive limit (i.e., when the probabilities of scattering to any angles are equal).

As it was shown later (see, e.g.,  review articles [\onlinecite{okulov79,falkovsky83}]), 
in reality the probability of scattering to
a certain angle strongly depends on the direction of momentum of incoming electron,
so that the scattering at small angle can be almost specular,
whereas it is rather diffusive for electrons incoming perpendicular to the surface.
Hence, the results of the calculation based on kinetic equation approach \cite{okulov79,falkovsky83}
has been compared to the results of approximation of constant parameter $p$ to
show that the specular parameter is not a constant, and the main contribution to conductivity
is related to most sliding electrons.

In this work we use essentially the same kinetic equation approach for the case of
two-dimensional graphene. Since graphene is the two-dimensional crystal, there is no
scattering from 2D surface as in thin films, and only the edge scattering is essential.
Thus, the direct comparison of the surface scattering in thin films and in graphene does
not make much sense. Nevertheless, we found for not too narrow graphene ribbon
that its conductivity can be presented as $\sigma =\sigma _0(1-Q\ell /L)$, with
$Q$ depending on the edge type and on the incoming angle (described by the
parameter $\xi $ (see Eqs. (29) and (30)). Note that both solutions (29),(30) are
valid only for $Q\ell /L\ll 1$. This is quite similar to the results for thin films and wires
with $\ell /L\ll 1$, \cite{falkovsky83}
where $L$ is the thickness or diameter of the sample. Here $Q$
substitutes the specular parameter $p$ and includes integration over all incoming
angles.

It should be stressed that the key point in the kinetic equation method relating the distribution functions
of incoming and outgoing
electrons is the probability of electron scattering at the surface. As we found, in the case of graphene this
probability is quite different for different types of the edges due to different boundary conditions for
the wave functions. In the case of zigzag boundary, one component of the wave function goes to zero
at the surface. As a result, the matrix element for the surface scattering at zigzag boundary has effectively
the same form as in conventional metal -- it is proportional to $k_x$, i.e., it is small for sliding electrons
(see Eqs. (19) and (20)).
In contrary, there is no such smallness for the Berry-Mondragon boundary.

Our calculations in the ballistic regime, $\ell /L\gg 1$ show that in the case of zigzag boundary, the relaxation time is formally
divergent for $k_x=0$. Namely, if $\xi \ll 1$, we get $\tau _z\sim k_x^{-2}$ (see Eq.~(37)). Correspondingly, by
using (41) we obtain $\sigma _z\sim L/k_{x,min}\sim L^2$. Note that the corresponding result is
$\sigma \sim (L/\ell )^{1/2}$ for thin films and $\sigma \sim (L/\ell )\log (\ell /L)$ for thin wires.

When the correction to relaxation time is mostly due to the scattering from curved edges
we found $\Delta \sigma \sim \ell /L$ with the coefficient depending on the variation of ribbon width.
For thin (ballistic) curved ribbon, $L/\ell \to 0$, we found $\sigma \sim L/k_Fa_\alpha ^2$
(where $a_\alpha $ is the characteristic length of fluctuations), see. Eq.~(55).

Note that there is no problem with sliding electrons for the case of ballistic disc because
the Berry-Mondragon boundary conditions for graphene disc lead to a constant electron relaxation time, with
$\tau \sim R$ (see Eq.~(94)).

\section{Conclusions}

We have considered different models of the boundary cnditions at the graphene edge to
calculate the electron relaxation time and conductance in graphene nanoribbons.
We have found that in the case of zigzag boundary the effect of edge scattering is
very strong. Similar to the surface scattering of electrons in conventional metals,
sliding electrons do not scatter from the zigzag edge. Thus, the edge scattering is
not effective for the nanoribbon with zigzag edges. In the case of Berry-Mondragon boundary,
the edge scattering can be the leading mechanism of electron scattering
determining the conductance of ballistic ribbons.

\section*{Acknowledgements}

This work is supported by the National Science Center in Poland as a research project in years
2011 -- 2014, by the Polish National Center of Research and Development under
Grant No.~UMO-2011/01/N/ST3/00394, and by the EC under the Graphene Flagship (contract no. CNECT-ICT-604391).

\end{document}